\documentclass{moriond}

\usepackage[symbol]{footmisc}
\usepackage[T1]{fontenc}

\def\be{\begin{equation}}
\def\ee{\end{equation}}
\def\bea{\begin{eqnarray}}
\def\eea{\end{eqnarray}}

\newcommand{\Photo}{}

\begin{document}
\vspace*{2cm}
\title{EXTERNAL ENVIRONMENTAL NOISE INFLUENCES \\ ON VIRGO DURING O3}

\author{N.Arnaud$^{1,5}$, R.DeRosa$^{2}$, F.DiRenzo$^{3,4}$, I.Fiori$^{5(*)}$, C.Giunchi$^{6}$, K.Janssens$^{7}$,\\ A.Longo$^{8}$, M.Olivieri$^{9}$, F.Paoletti$^{4,5}$, P.Ruggi$^{5}$, M.C.Tringali$^{5}$}
\address{
$^{1}$ Université Paris-Saclay, CNRS/IN2P3, IJCLab, 91405 Orsay, France\\
$^{2}$   Università Federico II and Istituto Nazionale di Fisica Nucleare, Sez. di Napoli, I-80126 Napoli, Italy\\
$^{3}$   Università di Pisa, Dipartimento di Fisica, I-56127 Pisa, Italy\\
$^{4}$   Istituto Nazionale di Fisica Nucleare, Sezione di Pisa, I-56127 Pisa, Italy\\
$^{5}$   European Gravitational Observatory (EGO), I-56021 Cascina, Pisa, Italy\\
$^{6}$  Istituto Nazionale di Geofisica e Vulcanologia, Sezione di Pisa, I-56125 Pisa, Italy\\
$^{7}$   Universiteit Antwerpen, Prinsstraat 13, 2000 Antwerpen, Belgium\\
$^{8}$  Istituto Nazionale di Fisica Nucleare, Sezione di Roma Tre, I-00146 Roma, Italy\\
$^{9}$  Istituto Nazionale di Geofisica e Vulcanologia, Sezione di Bologna, I-40128 Bologna, Italy\\
$^{(*)}${\rm Corresponding author,} irene.fiori@ego-gw.it
}

\maketitle\abstracts{
Sources of geophysical noise, such as wind, sea waves and earthquakes, can have an impact on gravitational wave interferometers causing sensitivity worsening and gaps in data taking.
During the 1-year long O3 run (April $1^{st}$ 2019 to  March $27$ 2020), the Virgo Collaboration collected a statistically significant dataset to study the response of the detector to a variety of environmental conditions. We used these data to correlate environmental parameters to global detector performance, such as observation range,  duty cycle and control losses. Where possible, we identified weaknesses in the detector and 
we elaborated strategies to improve Virgo robustness against external disturbances for the next run O4, planned to start in summer 2022. 
In this article we present preliminary results of this study.
}

\section{Introduction}
\label{sec:intro}
Advanced Virgo~\cite{AdV} is a second generation laser interferometer (ITF) for detecting gravitational waves (GW) located in Italy near Pisa. Each of the two 3~km long arms is a resonant Fabry-Perot cavity. The differential arm motion (DARM) produced by a GW changes the interference pattern of the laser beams onto the output photodiode. 
In the nominal working condition, or controlled state, 
a force is applied through coil-magnet actuators to the seismic isolation systems~\cite{SAT} of input and end mirror test masses in order to keep accurately at zero the differential arm length. 
The GW signal is extracted from the output photodiode. 

On 27 March 2020 the two LIGO, Virgo and GEO600 interferometers concluded 361 days of joined data taking, named the third observing run, O3. During O3, Virgo achieved a {\it BNS range}~\footnote{The binary neutron stars (BNS) range is a figure of merit of GW detectors sensitivity. BNS range is quoted as the average distance to which the signal generated by the coalescence of a system of two neutron stars with mass of 1.4M$_\odot$ could be detected with SNR of 8. 
\label{BNSrange}}
of $45 - 55$ Mpc, operated in {\it observing mode}~\footnote{The observing mode (or science mode) condition occurs when the interferometer is set in the nominal controlled state and no voluntary disturbing action (such as occasional tuning operations, periodic detector calibrations, weekly detector maintenance) is performed. Only data acquired in observing mode are used for GW signals searches.}
for $75\%$ of time, and lost the controlled state 
about 600 times. 
Periods of bad weather, corresponding to increased sea-wave activity and wind speed, were generally associated with increased noise in the GW signal below 100~Hz and some difficulty in maintaining the interferometer in the controlled state, resulting in reduced {\it duty cycle}~\footnote{The duty cycle is the fraction of the time the detector is taking data in observing mode.
\label{DutyCycle}}. 
We analyzed the large statistical sample collected during O3 to identify correlations between ITF figures of merit (BNS range, duty cycle and control losses) and environmental parameters (wind speed and ground seismicity) 
with the aim of understanding the Virgo interferometer behavior in different external environmental conditions and propose strategies to improve the detector robustness for the future O4 run, currently foreseen to begin in summer 2022. We present preliminary results of this study. A comprehensive dissertation will be the subject of a forthcoming publication.

\section{Characteristics of Virgo site ground seismicity and wind} \label{sec:characteristics}

An extended network of sensors monitors the Virgo physical environment~\cite{ENV_O3}. The seismic wave-field in the frequency range from 0.03~Hz to 50~Hz is monitored by three velocimeters Guralp 40T-60s, one for each experimental building located at the ITF vertexes. Velocimeters are deployed at the deep basement on which the seismic suspensions of the input and end mirror test masses rest. 
Virgo seismic wave-field is the sum of several sources whose contribution dominates in a specific frequency band. In the absence of earthquakes,  the next largest ground velocity noise is due to sea waves interaction with the shore. 
This occurs in the range between 0.1 and 1~Hz. 
The prominent frequency peak typically lays between 0.3~Hz and 0.4~Hz~\cite{SeismVirgo,KOLEY}. 
During O3, the 128s-averaged root mean square (RMS) value of vertical ground velocity in the 0.1~Hz to 1~Hz band ranged from $10^{-7}$~m/s to $10^{-5}$~m/s, while $10 \%$ of the time it exceeded  $4 \cdot 10^{-6}$~m/s. 
Ground seism motion between 1~Hz and~10 Hz has a marked daily cycle with drops during nights and holidays. The dominant sources are intermittent wave bursts from vehicle traffic on 1~km distant elevated motorways~\cite{SeismVirgo,KOLEY}. 
We measure a reduction of about $20\%$ of RMS noise in this band during the Covid-19 pandemic lockdown in spring 2020. 

A meteorological station (Davis Vantage-Pro-2) is located on the roof of a building close to the Virgo central experimental area, at about 10~m height from ground. It monitors wind speed and direction and other weather parameters 
at the rate of 0.4~Hz. Most of the time wind follows a 24-hour cycle, with minimum speed around mid-nights and maximum speed in early afternoons.
During O3 the average wind speed was 10~km/h, while it was larger than 20~km/h for $10\%$ of the time. Wind gusts exceeding 60~km/h occurred for $0.01\%$ of the time. Prevailing winds blew from east during $35\%$ of O3. 
Wind activity is associated with an increase of tilt noise of experimental buildings floor. 
The same effect was measured at LIGO and interpreted as due to the action of wind on experimental buildings walls~\cite{LIGO_O3}. 

\section{Wind influences}
\label{sec:wind}
\begin{figure}[htb!]
\begin{minipage}{0.49\linewidth}
\centerline{\includegraphics[width=0.9\linewidth]{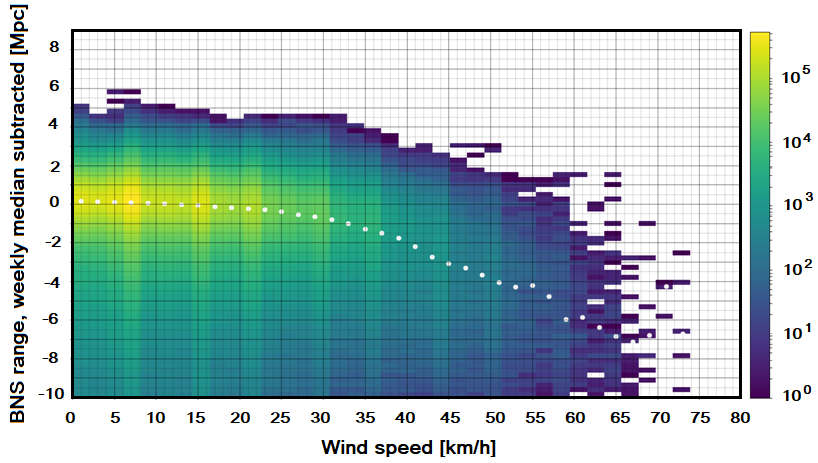}}
\end{minipage}
\hfill
\begin{minipage}{0.49\linewidth}
\centerline{\includegraphics[width=0.9\linewidth]{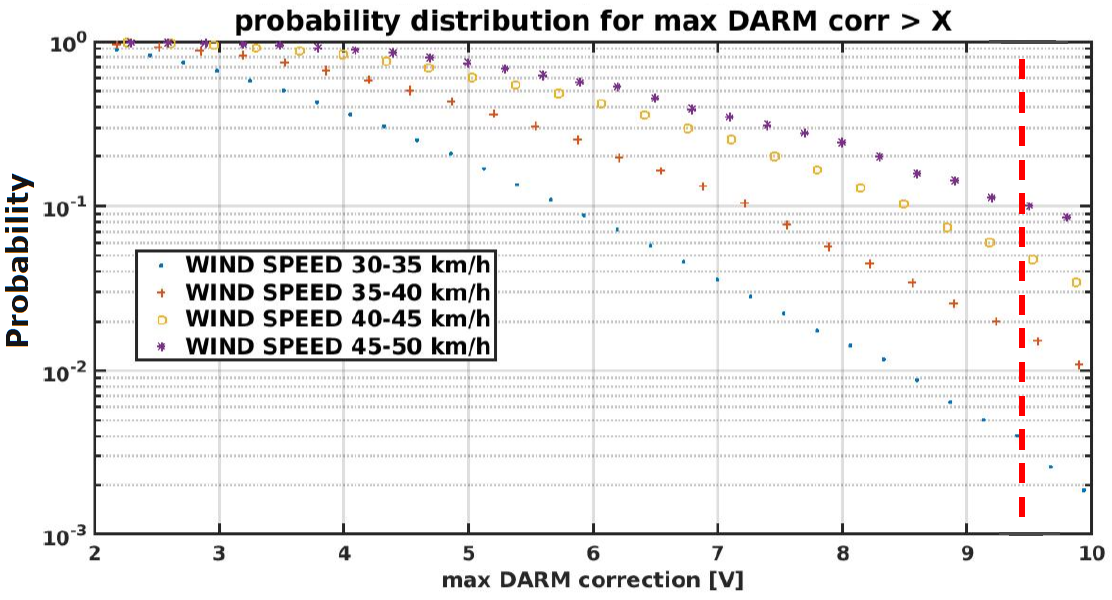}}
\end{minipage}
\caption[]{(left) Virgo BNS range deviation from weekly median is plotted as a function of wind speed, throughout O3 observing time. White dots are the average value computed in each wind speed bin. (right) Cumulative distribution of maximum control correction signal applied to the marionette stage of end mirror suspensions, $C_{max}$, in 4 different wind speed ranges between 30 and 50~km/h. Wind speed is averaged over 30-second long segments, and $C_{max}$ is the maximum correction within the same 30-second segment.
The red dashed line marks 9.5V: when $|C_{max}|$ reaches this value actuators saturate and the controlled state is always lost in a few seconds.}
\label{fig:1}
\end{figure}

A benchmark of interferometer sensitivity performance is the BNS range. It is a weighted average of the detector sensitivity curve, to which low frequency bins in the range from 30~Hz to 150~Hz contribute the most. Thorough O3, BNS range varied following an overall improving trend. 
Most variations occurred because of known reasons: occasional tuning of ITF global alignment~\cite{ISC_O3}, noise mitigations~\cite{ENV_O3}, technical improvements~\cite{ETALON}. 
A better figure to reveal residual unknown influences affecting ITF low frequency sensitivity is the BNS range deviation from the weekly median. 
In Figure~\ref{fig:1}(left), this quantity 
is plotted as a function of wind speed. We observe a degradation of detector sensitivity performance for wind speeds of 25~km/h and above.

Wind also had an impact on the robustness and average duration of the ITF controlled state. Our study evidences the correlation between wind speed and correction force applied to the suspension coil-magnet actuators to keep the arm optical cavities on resonance.
The larger the wind speed the larger the correction signal $C(t)$, up to the point that the maximum allowed correction signal ($|C(t)| = 10$~V) is reached and the controlled state is lost. Figure~\ref{fig:1}(right) illustrates this study. 
We found out that $14\%$ of O3 control losses were due to saturation of these actuators. A figure of merit we computed is the average remaining duration of the ITF controlled state after a given wind speed velocity is measured. As shown in Figure~\ref{fig:2}(left), this quantity rapidly decreases for maximum wind speeds above 20~km/h. 
A mitigation strategy was available, consisting in moving the control force from the end to the input marionette suspension stage actuators. Those actuators were driven by a different electronic board, providing 4 times more force. However, being these actuators more noisy and their calibration less accurate, it was decided to use them only in extreme environmental conditions, but out of the observing mode. 
For O4, it is foreseen to have a low noise actuation in use both for the input and the end suspension marionettes, having a factor $\sqrt{2}$ more force with the same total noise.

\section{Sea influences}
\label{sec:sea}
\begin{figure}[htb!]
\begin{minipage}{0.49\linewidth}
\centerline{\includegraphics[width=0.9\linewidth]{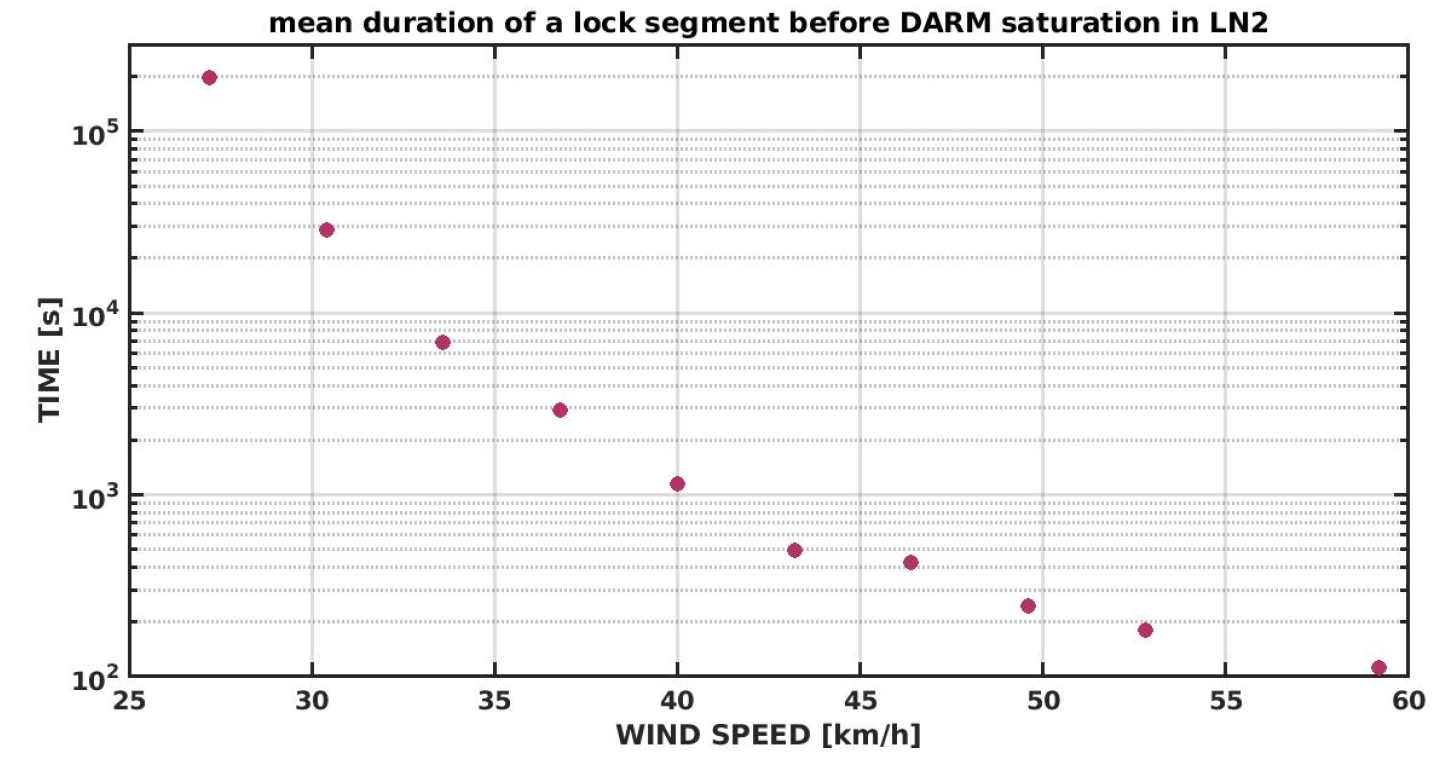}}
\end{minipage}
\hfill
\begin{minipage}{0.49\linewidth}
\centerline{\includegraphics[width=0.9\linewidth]{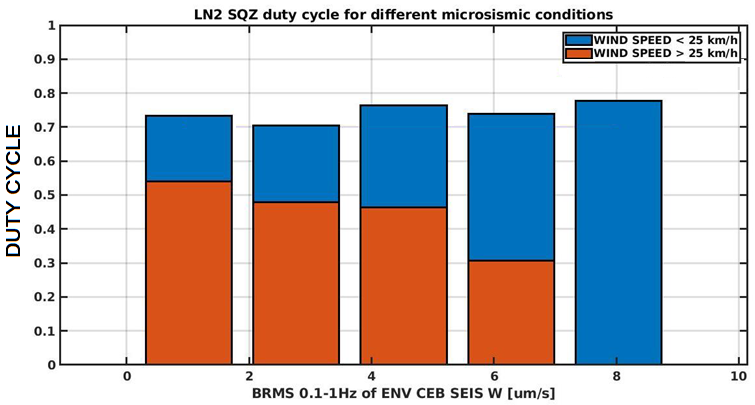}}
\end{minipage}
\caption[]{(left) 
Estimated remaining duration of Virgo controlled operation ($T_{lock}$) after a given 30-second averaged wind speed is measured. This quantity is computed as: $T_{lock}=30$s$/P$($C_{max} > 9.5$V). Where, $P$ is the probability that the correction exceeds 9.5~V.
(right) Virgo duty cycle is computed as a function of RMS ground seismicity in the 0.1~Hz to 1~Hz frequency band, for two complementary data samples: low wind speed time periods (blue) and high wind speed time periods (red).}
\label{fig:2}
\end{figure}

Because of the wind action on sea surface, high winds and sea induced microseism often occur together or with just a few hours delay, and it is difficult to select long time laps in which just one of the two dominates. To disentangle effects of wind and sea on Virgo we followed a statistical approach. We divided the O3 dataset in two subsets according to wind speed ($v$): a {\it low-wind} subset ($v < 25$~km/h) and {\it high-wind} subset ($v \ge 25$~km/h). For both samples we computed the detector duty cycle as a function of RMS ground seismicity in the frequency range 0.1~Hz to 1~Hz. The result is shown in Figure~\ref{fig:2}(right). In the high-wind sample the duty cycle reduces with increasing ground seismicity, while in the low-wind sample it is essentially independent on it. In other words, the control state is not affected by microseismic activity as long as wind speed is low. Our analysis shows that the ITF control scheme adopted during O3~\cite{ISC_O3,SAT_O3} is quite robust against microseismic activity.

On the other hand, Virgo sensitivity was typically bad when sea microseism was high. 
The main cause of sensitivity degradation was low frequency noise up to ${\sim} 100$~Hz associated to scattered light effects inside the interferometer~\cite{SL_WAS,ENV_O3}. 
Tiny stray light beams may re-couple to the interferometer main laser mode adding the phase noise they acquired along the path. 
In Virgo, one source of scattered light is the suspended auxiliary optics, placed for example in transmission of the two arms. The Suspended West End Bench (SWEB) was identified as the major culprit, also by means of adaptive search algorithms~\cite{SL_LONGO}.
Despite a force is exerted on the bench suspension top stage to minimize the relative velocity between the bench and the end mirror, the SWEB was moving too much at the frequency of the microseismic peak. 
This occurred due to a defect in the suspension inertial damping control, which was identified and cured in preparation for O4.

\section{Earthquakes}
\label{sec:earthquakes}
Another issue we started investigating is noise induced by earthquakes (EQ). During O3, 54 earthquakes were either strong enough or close enough to cause control losses.
EQ alerts were provided by the {\it Seismon}~\cite{Seismon} software tool developed by LIGO and adapted to Virgo. 
If the warning was significant enough (based on a rough magnitude-distance cut) and it was received early enough (i.e. before the seismic waves) the ITF control system was manually switched to a more robust configuration.
Seismon relies on the Earthquake Early Warning (EEW) data streaming provided by the United States Geological Survey and a seismic model to predict waves arrival time and amplitude at the Virgo site.
At this preliminary stage we have identified strategies to improve the effectiveness of the warning system. One way is to improve the coverage to European and Mediterranean events, which are most significant for Virgo. To this end, we plan to integrate in Seismon the EEW data stream from 
{\it EarlyEst}~\cite{EarlyEst},
a lightweight software package for rapid global scale earthquake monitoring which reads data 
also from seismic networks in the Mediterranean region.
The new system performance can be
estimated a priori using O3 Virgo
data. 
In case of close earthquakes (few hundreds km from Virgo) the warning might not arrive fast enough to be received and processed. The solution explored by the project ASPIS~\cite{ASPIS} 
is to implement a dedicated network of seismometers  located at 100 km from Virgo to intercept seismic waves and adopting a fast EQ identification algorithm. A complementary approach is to 
improve the robustness of the Virgo controls. The plan is using close seismic shakes collected during O3  
to study the response of Virgo and identify sensible degrees of freedom of the interferometer.


\end{document}